\documentclass[a4paper,11pt]{article}
\usepackage{graphicx,amssymb,amsmath,amsfonts,epsfig}
\usepackage[usenames,dvipsnames]{color}
\usepackage{subfigure}
\title{Entropy Product Function and Central charges in NUT Geometry}
\author{Parthapratim Pradhan\footnote{pppradhan77@gmail.com}\\ 
\textit{Department of Physics}\\
\textit{Hiralal Mazumdar Memorial College For Women}\\
{Dakshineswar, Kolkata-700035, India}}
\date{}

\begin{document}

\maketitle

\def \be {\begin{equation}}
\def \ee {\end{equation}}
\def \ba {\begin{array}}
\def \ea {\end{array}}
\def \bea{\begin{eqnarray}}
\def \eea{\end{eqnarray}}

\def \a {\alpha}
\def \b {\beta}
\def \g {\gamma}
\def \G {\Gamma}
\def \d {\delta}
\def \D {\Delta}
\def \e {\epsilon}
\def \ve {\varepsilon}
\def \m {\mu}
\def \n {\nu}
\def \k {\kappa}
\def \l {\lambda}
\def \L {\Lambda}
\def \s {\sigma}
\def \S {\Sigma}
\def \r {\rho}
\def \o {\omega}
\def \O {\Omega}
\def \th {\theta}
\def \Th {\Theta}
\def \t {\tau}
\def \z {\zeta}

\def \p {\partial}
\def \f {\frac}
\def \na {\nabla}
\def \da {\Box}
\def \nn {\nonumber}
\def \scl {\ell}
\def \ma {\mathcal}
\def \mb {\mathbb}
\def \mc {\mathcal}

\def \lt {\left}
\def \rt {\right}
\def \sl {\slashed}
\def \la {\leftarrow}
\def \ra {\rightarrow}
\def \lra {\leftrightarrow}
\def \sr {\sqrt}
\def \td {\tilde}
\def \hs {\hspace}
\def \pp {\propto}
\def \inf {\infty}
\def \dd {\mathrm{d}}

\begin{abstract}
We define an \emph{entropy product function}~(EPF) for Taub-Newman-Unti-Tamburino~(TNUT) 
black hole~(BH) following the prescription suggested by Wu et al.~\cite{wu}
~[PRD 100, 101501(R) (2019)]. 
The prescription argues that a generic four-dimensional TNUT spacetime might be expressed 
in terms of three or four different types of thermodynamic hairs. They can be  
defined as the Komar mass~($M=m$), the angular momentum~($J_{n}=mn$), the gravitomagnetic 
charge ($N=n$), the dual~(magnetic) mass $(\tilde{M}=n)$. Taking this prescription and 
using the \emph{EPF}, we derive  the \emph{central charges} of dual CFT~(conformal field theory) 
via  Cardy's formula. 
Remarkably, we \emph{find} that for TNUT BH there exists a relation between the 
\emph{central charges and EPF} as $c=6\left(\frac{\partial {\cal F}}{\partial {\cal N}_{i}}\right)$, 
where ${\cal F}$ is EPF and ${\cal N}_{i}$ is one of the integer-valued 
charges i.e. the NUT charges~($N$) or any new conserved charges~($J_{N}$). 
We reverify these results by calculating the exact values of different thermodynamic parameters. 
We define the EPF~${\cal F}$ from the first law of thermodynamics of both 
horizons. Moreover, we write the first laws of both the horizons for left-moving and right-moving sectors.
Introducing the B\'{e}zout's identity, we show that for TNUT BH one can generate 
more holographic descriptions described by a pair of integers $(a,b)$. More holographic pictures have 
a great significance towards understanding the holographic nature of quantum gravity. 
Furthermore, using the \emph{EPF} we derive the central charges for Reissner-Nordstr\"{o}m-NUT~(RNNUT) BH, 
Kerr-Taub-NUT~(KNUT) BH and Kerr-Newman-NUT~(KNNUT) BH. Finally we  prove that they are 
equal in both sectors provided that the EPF is mass-independent~(or universal).
\end{abstract}

%\keywords{Entropy product function, TNUT BH, KNUT BH, KNNUT BH }

\section{Introduction}
The Kerr/CFT correspondence~\cite{guica} is a great discovery in BH physics 
to understand the entropy of a BH. It is the macroscopic entropies of BHs which 
has been reproduced exactly from the microscopic states counting in the dual 2D CFT 
via Cardy formula. The fact started for many years that the microscopic origin of 
the  BH entropy could be holographically encoded in a 2D CFT, particularly after 
successfully  counting of the Bekenstein-Hawking entropies of a extremal BH in 
string theory~\cite{vafa}. 

One  aspect is that the greybody factors for rotating 
black holes in four dimensions were derived by Cvetic and Larsen~\cite{cv97} to 
study the Hawking radiation. The string 
excitations in both sectors i.e. left and right moving sectors generate
two distinct contributions of entropy. Similarly, the emission spectrum of 
the string is identified by left moving and right moving which is independent 
of temperatures. Consequently the appearance of two independent sets of 
thermodynamic variables is related to the presence of event horizon and 
Cauchy horizons. For instance, the two contributions to the entropy are 
proportional to the sum and difference of the Cauchy and event horizon 
area, respectively. This geometic interpretation of the thermodynamic 
variables gives access to features of the underlying microscopic 
theory.

On the other hand the solution of wave equation for most 
general spacetime background is derived to compute greybody factors in 
four and five dimensions~\cite{cvf97}. Interestingly the wave equation has 
an exact symmetry that interchanges the Cauchy~(or inner) horizon and 
event~(or outer) horizons. Importantly, this symmetry has a discrete 
in nature. The modes in the vicinity of the event horizon give rise 
to the Hawking radiation having Hawking temperature 
$T_{H}=\frac{T_{L}+T_{R}}{2T_{R}T_{L}}$. Similarly,  
from the modes in the vicinity of the Cauchy horizon give rise to a 
characteristic temperature $T_{C}=\frac{T_{L}-T_{R}}{2T_{R}T_{L}}$.
The temperatures $T_{R}$ and $T_{L}$ that appear in these formulas 
agree precisely with those that follow from thermodynamics.

One of the interesting feature in the holographic description of  KN class of BHs is 
that the central charges of the dual CFT are independent of the mass of BHs.  The 
attribute could be related to the fact that the entropy product of ${\cal H}^{\pm}$ 
of these BHs are also mass-independent.

In the context of general relativity, it has been examined that for a four-dimensional 
Kerr-Newman BH~\cite{ansorg09}, the entropy product of ${\cal H}^{\pm}$ i.e.  
\begin{eqnarray}
{\cal S}_{+} {\cal S}_{-} &=& 4\pi^2\left(J^2+\frac{Q^4}{4} \right) 
\end{eqnarray}
is mass-independent. On the other hand in the context of string theory, it has been 
argued that for a BPS class of BHs, the entropy product of ${\cal H}^{\pm}$  should 
be quantized  as~\cite{cvetic10}
\begin{eqnarray}
{\cal S}_{+} {\cal S}_{-}  &=& (2\pi)^2 \left[\sqrt{N_{1}}+\sqrt{N_{2}}\right]
\left[\sqrt{N_{1}}-\sqrt{N_{2}}\right]
= (2\pi)^2 N , \,\, N\in {\mathbb{N}}, N_{1}\in {\mathbb{N}}, N_{2} \in {\mathbb{N}} 
~.\label{tnl}
\end{eqnarray}

It is an well-established fact that the thermodynamics method
~\footnote{The conventional way to determine the cental charges of dual CFT by 
using asymptotic symmetry group~(ASG) of near-horizon geometry of extremal BH in either 
BBC~(Branich-Brandt-Compere) formalism`\cite{guica} or equivalently the stretched horizon 
formalism~\cite{carlip12}} is a powerful mechanism to determine 
the dual CFT in Kerr/CFT correspondence~\cite{chen12}. In case of Kerr/CFT correspondence it has 
been proved that more universal information of the dual CFT including the dual temperatures, 
and the central charges of both left-moving sectors and right-moving sectors  are fully 
encoded in the thermodynamics of the event~(or outer) horizon~(EH)~(${\cal H}^{+}$) 
and Cauchy~(or inner) horizon~(CH)~(${\cal H}^{-}$) of the Kerr-Newman~(KN) BH. 
Also it has been suggested in~\cite{chen12} that the thermodynamics method is 
universal to determine the dual holographic picture of BH.

For KN BH~\cite{chen12} it has been explicitly proved that the first laws of BH 
thermodynamics satisfied for EH and CH horizons as 
\begin{eqnarray}
dM &=& \pm {T}_{\pm}\, d{\cal S}_{\pm} + \Omega_{\pm}\, dJ +\Phi_{\pm}\,dQ ~. \label{ee1.1}
\end{eqnarray}
where $T_{\pm}$, $\Omega_{\pm}$ and $\Phi_{\pm}$ represents the Hawking temperature, the angular 
velocity and the electric potential of ${\cal H}^{\pm}$ calculated on the horizons for KN BH.
Under the exchange of $r_{+}\leftrightarrow r_{-}$ and using the symmetry of outer~(inner) horizons
$r_{\pm}$, one would get the Hawking temperature of the inner horizon 
\begin{eqnarray}
 T_{-} &=&- T_{+}\Big|_{r_{+}\leftrightarrow r_{-}}
\end{eqnarray}
while the other quantites are changed under the symmetry of $r_{\pm}$ as 
\begin{eqnarray}
\mathcal{S}_{-} &=& \mathcal{ S}_{+}\Big|_{r_{+}\leftrightarrow r_{-}},\,\,
\Omega_{-} =\Omega_{+}\Big|_{r_{+}\leftrightarrow r_{-}},\,\,
\Phi_{-} = \Phi_{+}\Big|_{r_{+}\leftrightarrow r_{-}}
\end{eqnarray}
An another universal relation exists for KN BH 
\begin{eqnarray}
T_{+}{\cal S}_{+}= T_{-}{\cal S}_{-}  ~~\label{eq4.3}
\end{eqnarray}    
which must says that the entropy product is \emph{mass independent}. 

In recent times, it has been proposed that~\cite{wu} a generic four dimensional 
TNUT spacetime has four thermodynamical hairs. They could be defined as the 
Komar mass ($M=m$), the angular momentum~($J_{n}=mn$), the gravitomagnetic 
charge~($N=n$), and or the dual~(magnetic) mass~$(\tilde{M}=n)$. Using this formalism, 
\emph{in this work} we prove that there exists a relation between \emph{EPF and 
the central charges of dual CFT for TNUT BH}. First, we define the EPF from the first law of
BH thermodynamics for both the horizons  and then using this function we determine the 
central charges of dual CFT. Moreover, we calculate the thermodynamic parameters in 
left-moving sectors and right-moving sectors. Furthermore, we compute the first 
laws of thermodynamics in left-moving sectors and right-moving sectors. Finally, we 
examine the holographic picture of TNUT BH. Since the BH carries NUT parameter~($N$)
~\footnote{NUT charge should be considered as a thermodynamic variable since it is 
independently varied in the full cohomogeneity of first law in BH thermodynamics.}
and new conserved charges ($J_{N}$) hence there exists two elementary holographic 
pictures, which we call the $N$ picture and the $J_{N}$ picture.

Like four dimensional dyonic RN BH here in the context of \emph{NUT class of BHs}  
we prove that there exists a relation between the EPF~(${\cal F}$) 
and the central charges~($c$) as 
\begin{eqnarray}
c^{i}_{L,R} = 6\left(\frac{\partial {\cal F}}{\partial {\cal N}_{i}}\right) ~\label{ee2.1}
\end{eqnarray}
where  ${\cal N}_{i}$ is one of the integer-valued  charges appearing in the first laws,  
and should be angular momentum, or another conserved charges. For instance i.e. for  RN BH, the 
EPF~\cite{chen12} is defined as 
\begin{eqnarray}
\mathcal{F} = \frac{{\cal S}_{+}{\cal S}_{-}}{4\pi^2}= \frac{Q^4}{4}. % ~\label{ee3.1}
\end{eqnarray}
Since the EPF is mass-independent hence $T_{+}{\cal S}_{+} = T_{-}{\cal S}_{-}$. 
Therfore one can  derive the central charges for $Q$ picture by using Eq.~(\ref{ee2.1}) as 
\begin{eqnarray}
c_{L}^{Q} &=& 6Q^3 \\
c_{R}^{Q} &=& 6Q^3  ~\label{app1.2}
\end{eqnarray}
They are equal for left moving and right-moving sectors that means~$c_{L}^{Q}=c_{R}^{Q}$.
The central charges are equal means that the EPF is universal~(mass-independent). 
This result is completely agreement with the central charges  derived by asymptotic symmetry group (ASG) 
analysis~\cite{ghodsi10}. Where it was proved by uplifting the four-dimensional compactification solution 
to a five-dimensional solution.

In the literature~\cite{jetp,epl,mpla}, it has been shown that the area~(or entropy) 
products in NUT geometry i.e. TNUT BH,  KTNUT BH and KNTNUT 
have ``mass-dependent'' characteristics. 
Secondly, the first laws of thermodynamics in the left moving sectors and the right moving sectors 
do not satisfied like Eq.~(\ref{e1.1}). In that situation,  the EPF is
mass-dependent that means  
\begin{eqnarray}
T_{+}{\cal S}_{+} \neq  T_{-}{\cal S}_{-} ~~\label{e1.2}
\end{eqnarray}  
Moreover, we cannot read the information of dual CFT that means we 
cannot derived the exact left-moving and right-moving temperatures of dual CFT. 
Furthermore, the  central charges of left-moving sectors and right-moving sectors are 
not equal i.e. 
\begin{eqnarray}
c_{L}\neq c_{R} 
\end{eqnarray}
However incorporating the fromalism stated in~\cite{wu}, we should be able to defined
the EPF for NUT class of BHs because the entropy product of 
${\cal H}^{\pm}$ is universal i.e. mass-independent~\cite{plb20}. In this work we will 
consider the case $J_{N}=MN$ and the case $J_{N}\neq MN$ is already been discussed in 
the following work~\cite{jetp,mpla,epl}. Moreover in this work we derive the central 
charges from entopy product function and compared it with the result obtained from the 
ASG analysis for NUT class of BHs.

There are two ways one can  compute the central charges. The first one is that 
ASG of near horizon geometry of extremal BH and the second one is that via EPF 
method i.e. thermodynamics method. In this work we will particularly emphasize 
on EPF method. Also we will see that \emph{EPF method is more covenient to derive the 
central charges than the ASG method}.  

The outline of the Letter is as follows.  In the next section~~(\ref{ntn}), we derive the 
relation between EPF and central charges for NUT class of BH i.e. TNUT 
BH. In Sec.~(\ref{etn}), we derive the exact calculations of central charges by direct calculation.  
In Sec.~(\ref{rnn}), we derive the central charges for RNNUT BH by using EPF. 
In Sec.~(\ref{knn}), we derive the central charges for KNUT BH by using EPF.
In Sec.~(\ref{knnn}), we derive the central charges for KNNUT BH by using EPF.
In Sec.~(\ref{con}),  we have given our conclusions. In Appendix-A~(Sec. 8), we evaluate 
the central charges from EPF for RN BH, Kerr BH and KN BH. In Appendix-B~(Sec. 9), we 
provide the calculation of central charges by using near horizon geometry of extremal
KN BH.

\section{\label{ntn} TNUT BH and EPF}
The Lorentzian TNUT BH~\cite{mkg,miller} is a solution of the Einstein equation. 
The metric is defined as
\begin{eqnarray}
ds^2 &=& -{\cal A} \, \left(dt+2n\cos\theta d\phi\right)^2+ 
\frac{dr^2}{{\cal A}}+\left(r^2+n^2\right) \left(d\theta^2
+\sin^2\theta d\phi^2 \right) ~,\label{tn1}
\end{eqnarray}
where the function ${\cal A}$ is
\begin{eqnarray}
 {\cal A} &=& \frac{1}{r^2+n^2} \left[r^2-n^2-2mr \right]
\end{eqnarray}
Under the new prescription~\cite{wu} for NUT class of BHs, the global 
conserved charges computed via conformal completion method~\cite{wu22} 
could be defined  as 
\begin{eqnarray}
\mbox{Komar mass}: \,\,m &=& M~ \nonumber\\
\mbox{Angular momentum}: m\,n &=& J_{n}~\nonumber\\
\mbox{Gravitomagnetic charge}:\,\, n &=& N \\
\mbox{Dual~(magnetic) mass}:\,\, \tilde{M} &=& n
~.\label{ptn}
\end{eqnarray}
It means that NUT charge is a thermodynamic multihair. 
It further implies that simultaneously it has both rotation-like 
and electromagnetic charge-like properties.

Taking the effect of Eq.~({\ref{ptn}}), the metric can be written as 
\begin{eqnarray}
ds^2 &=& -{\cal A} \, \left(dt+2N\,\cos\theta d\phi\right)^2+ 
\frac{dr^2}{{\cal A}}+\left(r^2+N^2\right)
\left(d\theta^2+\sin^2\theta d\phi^2 \right) ~,\label{tn}
\end{eqnarray}
where the function ${\cal A}$ is given by 
\begin{eqnarray}
{\cal A} &=& \frac{1}{r^2+N^2} \left(r^2-N^2-2Mr\right)
\end{eqnarray}
The BH horizons are at
\begin{eqnarray}
r_{\pm}= M \pm \sqrt{M^2+N^2}\,\, \mbox{and}\,\,  r_{+}> r_{-}
\end{eqnarray}
$r_{+}$ is called EH  and $r_{-}$  is called CH. Thus the 
entropy~\cite{plb20,grg21} of the BH is computed under the 
formalism stated in Eq.~(\ref{ptn})
\begin{eqnarray}
{\cal S}_{\pm} &=& 
%\frac{1}{4} \int^{2\pi}_0\int^\pi_0\sqrt{g_{\theta\theta}\,g_{\phi\phi}}{\mid }_{r=r_{\pm}} d\theta\, d\phi 
                2\pi\left[M^2+N^2 \pm \sqrt{M^4+J_{N}^2}\right]
~.\label{ctn}
\end{eqnarray}
One can compute the BH temperature by defining partition 
function of a well-defined microcanonical ensemble. The fact that a BH solution 
having a topological charge should be described by the BH thermodynamics. It was 
started by Gibbons and Hawking in 1977~\cite{Gibbons77}. They proposed that the 
partition function of a  canonical ensemble for BHs should be calculated with its 
Euclidean action in the form of the gravitional path integral. It is described by 
\begin{eqnarray}
\cal{Z} &=& e^{-\beta F} =\int D[g]e^{-\frac{I}{h}}\sim e^{-\frac{I}{h}}, 
\end{eqnarray}
where $I$ and $F$ is the Euclidean action and the free energy of the BH. 
The period $\beta$ of the Euclidean time is the inverse of the BH  
temperature i.e. $\beta=\frac{1}{T}$.
For Taub-NUT BH  and for both the horizons ${\cal H}^{\pm}$, 
$\frac{I_{\pm}}{\beta_{\pm}}=F_{\pm}=\frac{m}{2}$, $\beta_{\pm}=\frac{1}{T_{\pm}}$ 
is the interval of the time coordinate~\cite{Mann19}.

In our earlier work~\cite{grg21}, we proved that mass parameter can 
be expressed as a function of area~(or entropy), new conserved charges and NUT parameter 
for both the horizons i.e. $M=M({\cal S}_{\pm},J_{N}, N)$. Then the  first law of 
thermodynamics is satisfied for both the horizons  as 
\begin{eqnarray}
dM &=& T_{+} d{\cal S}_{+}+\omega_{+}dJ_{N}+\psi_{+} dN,~\label{eq4.0}\\
   &=& -T_{-} d{\cal S}_{-}+\omega_{-}dJ_{N}+\psi_{-} dN ~\label{eq4.1}
\end{eqnarray}
where
\begin{eqnarray}
T_{\pm} &=& \frac{r_{\pm}-M}{2\pi \left(r_{\pm}^2+N^2 \right)},\,\, 
\omega_{\pm}= \frac{N}{r_{\pm}^2+N^2},\,\,
\psi_{\pm}  = -\frac{2N\,r_{\pm}}{r_{\pm}^2+N^2}~\label{eq1.3}
\end{eqnarray}
Like KN BH, there exists a symmetry for NUT class of BHs between the following physical
quantites under exchange of two physical horizons:
\begin{eqnarray}
T_{-} &=& -T_{+}|_{r_{+}\leftrightarrow r_{-}} \\
{\cal S}_{-} &=& {\cal S}_{+}|_{r_{+}\leftrightarrow r_{-}}\\
\omega_{-} &=& \omega_{+}|_{r_{+}\leftrightarrow r_{-}}\\
\psi_{-} &=& \psi_{+}|_{r_{+}\leftrightarrow r_{-}} ~\label{eq4.2}
\end{eqnarray}
This indicates that if the first law of thermodynamics satisfied on the event horizon then it 
must be satisfied on the Cauchy horizon. Using Eq.~(\ref{eq4.0}) and Eq.~(\ref{eq4.1}) we find
\begin{eqnarray}
d\left({\cal S}_{+}{\cal S}_{-}\right) = 
\left(\frac{{\cal S}_{-}T_{-}-{\cal S}_{+}T_{+}}{T_{+}T_{-}}\right)\,dM
+\left(\frac{{\cal S}_{+}T_{+}\omega_{-}-{\cal S}_{-}T_{-}\omega_{+}}{T_{+}T_{-}}\right)\,dJ
+\left(\frac{{\cal S}_{+}T_{+}\psi_{-}-{\cal S}_{-}T_{-}\psi_{+}}{T_{+}T_{-}}\right)\,dQ \nonumber\\
~\label{e3}
\end{eqnarray}
It can be rewritten as 
\begin{eqnarray}
d\mathcal{F} = 
\left(\frac{{\cal S}_{-}T_{-}-{\cal S}_{+}T_{+}}{4\pi^2\,T_{+}T_{-}}\right)\,dM
+\left(\frac{{\cal S}_{+}T_{+}\omega_{-}-{\cal S}_{-}T_{-}\omega_{+}}{4\pi^2\,T_{+}T_{-}}\right)\,dJ
+\left(\frac{{\cal S}_{+}T_{+}\psi_{-}-{\cal S}_{-}T_{-}\psi_{+}}{4\pi^2,T_{+}T_{-}}\right)\,dQ~. \nonumber\\
\label{e1.1}
\end{eqnarray}
where
\begin{eqnarray}
\mathcal{F} &=& \frac{{\cal S}_{+}{\cal S}_{-}}{4\pi^2} ~\label{ee3.1}
\end{eqnarray}
This $\mathcal{F}$ is defined as \emph{EPF} for TNUT BH. We first defined the 
EPF from the first law of thermodynamics of both the horizons. Again we know for TNUT 
BH the entropy product of ${\cal H}^{\pm}$~\cite{plb20} is 
\begin{eqnarray}
\frac{{\cal S}_{+}{\cal S}_{-}}{4\pi^2} &=& J_{N}^2+N^4 ~\label{q4.4}
\end{eqnarray}
Thus 
\begin{eqnarray}
\mathcal{F} &=&  J_{N}^2+N^4 ~\label{e3.1}
\end{eqnarray}
where ${\cal F}$ is function of $J_{N}$ and $N$. The term \emph{EPF} was first
introduced in~\cite{chen13,chen13a} for investigation of the holographic pictures of 
four dimensional dyonic RN BH. The Eq.~(\ref{e1.1}) implies that $\mathcal{F}$ is 
independent of the mass parameter indicates that
\begin{eqnarray}
T_{+}{\cal S}_{+} &=&  T_{-}{\cal S}_{-}. ~~\label{ee1.2}
\end{eqnarray}
To derive the Smarr formula for TNUT BH first we have to consider the dimensions of 
the following thermodynamic parameters:
\begin{eqnarray}
[M]=[N]=L,\,\, [{\cal S}_{+}]=[{\cal S}_{-}]=[J_{N}]=L^2
\end{eqnarray}
Thus taking $({\cal S}_{+}, J_{N}, N)$ as independent parameters  and $M$ is the 
homogenous function of $({\cal S}_{+}^{\frac{1}{2}}, J_{N}^{\frac{1}{2}}, N)$ with 
degree 1. Then Euler's homogenous function theorem indicates that 
\begin{eqnarray}
M &=& 2{\cal S}_{+}\frac{\partial M}{\partial {\cal S}_{+}}+2J_{N}\frac{\partial M}{\partial J_{N}}
+N\frac{\partial M}{\partial N}.
\end{eqnarray}
Hence the Smarr formula is 
\begin{eqnarray} \label{e10}
M &=& 2\left(T_{+}{\cal S}_{+} +\omega_{+} J_{N}\right)+\Psi_+ N.
\end{eqnarray}
Similarly, the  Smarr formula for the Cauchy horizon is given by
\begin{eqnarray} \label{e11}
M &=& 2\left(-T_{-}{\cal S}_{-}+\omega_{-}J_{N}\right)+\Psi_{-} N.
\end{eqnarray}
Using the Smarr formulae for the event horizon and Cauchy horizons~(\ref{e10}) and (\ref{e11}), we can 
see that~(\ref{e3}) implies that
\begin{eqnarray} \label{e4}
\dd \ln(\mathcal{F}) &=& \f{(\o_--\o_+)\dd J_{N}+(\Psi_--\Psi_+)\dd N}{2\pi^2(\o_--\o_+)J_{N}
+\pi^2N(\Psi_--\Psi_+)}.
\end{eqnarray}
When $J_{N}\neq 0$ and $N\neq 0$, in general we can write
\begin{eqnarray}
\f{\o_--\o_+}{\Psi_--\Psi_+}&=& g(J_{N},N),
\end{eqnarray}
where $g(J_{N},N)$ is some unknown function of $(J_{N},N)$. Thus we have
\begin{eqnarray} \label{e1}
\dd \ln(\mathcal F)=\f{g\,\dd J_{N}+\dd N}{2\pi^2g\,J_{N}+N\pi^2}.
\end{eqnarray}
Consistency requires that
\begin{eqnarray}
\p_{N}\f{g}{2J_{N}g+N}=\p_{J_{N}}\f{1}{2J_{N}\,g+N},
\end{eqnarray}
which leads to
\begin{eqnarray}
2J_{N}\p_{J_{N}}\,g+N\p_N g=-g.
\end{eqnarray}
This means that $g$ is the homogeneous function of $(J_{N}^{\f{1}{2}},N)$ with degree $-1$, thus we may 
write $g(J_{N},N)=N^{-1}f(N^2/J_{N})$. Integrating (\ref{e1}) we get
\begin{eqnarray}
\mathcal{F} \pp J_{N}^2\, f(N^2/J_{N}),
\end{eqnarray}
where
\begin{eqnarray}
\ln f(x)=\int^x_0 \f{\dd y}{2f(y)+y}.
\end{eqnarray}
Hence the EPF $\mathcal{F}$ is the homogeneous function of 
$(J_{N},N^2)$ with degree 2, thus quasi-homogeneous function of  $(J_{N},N)$.

Following Refs.~\cite{cv97,cvf97}, we can define the thermodynamic parameters  in the 
left-moving and right-moving sectors for TNUT BH as
\begin{eqnarray}
T_{L,R} &=& \frac{T_{+}T_{-}}{T_{-}\pm T_{+}}, \nonumber\\
{\cal S}_{L,R} &=& \frac{{\cal S}_{+}\pm {\cal S}_{-}}{2},\nonumber\\
\omega_{L,R} &=& \frac{T_{-}\,\omega_{+}\pm T_{+}\,\omega_{-}}{2(T_{-}\pm T_{+})} \nonumber\\
\psi_{L,R}  &=& \frac{T_{-}\,\psi_{+}\pm T_{+}\,\psi_{-}}{2(T_{-}\pm T_{+})}
~\label{eq4.5}
\end{eqnarray}
It was demonstrated in Ref.~\cite{cv97,cvf97} the first law of thermodynamics for left-moving and 
right-moving sectors are satisfied separately in the context of rotating BH in four dimensions, 
and also there exists separate Smarr formulae for two sectors. 
Here, we \emph{show} that such type of relation exists for \emph{NUT class of BH} also. Thus in terms of 
left and right moving modes of dual CFT,  the  first law of BH thermodynamics can be rewritten as 
\begin{eqnarray}
\frac{dM}{2} &=& T_{L}\, d{\cal S}_{L}+\omega_{L}\, dJ_{N}+\psi_{L}\, dN   \nonumber \\
             &=& T_{R}\, d{\cal S}_{R}+\omega_{R}\, dJ_{N}+\psi_{R}\, dN  ~.\label{eq4.6}
\end{eqnarray} 
Now we shall keep $J_{N}$ as a invariant quantity and taking the perturbation of type 
$(dN, dJ_{N})=dN(1,0)$ then we should get from the first law 
\begin{eqnarray}
dN=\frac{T_{L}}{\psi_{R}-\psi_{L}} \, d{\cal S}_{L}-\frac{T_{R}}{\psi_{R}-\psi_{L}} \, d{\cal S}_{R} ~.\label{eq4.7}
\end{eqnarray}
From the above computations, we can read the information of the dual CFT. Also, we can predict two 
important facts:\\ 
(a) ${\cal S}_{L}$ and ${\cal S}_{R}$ in Eq.~(\ref{eq4.5}) are the exact  entropies of the left
and right moving sectors of dual CFT~\cite{cv96,cv97,cvf97,cvf09}.\\
(b) When we keep $J_{N}$ is fixed, we get the $N$ picture i.e the NUT picture. There exists the 
first law of thermodynamics
\begin{eqnarray}
dN &=& T_{L}^{N} d{\cal S}_{L} -T_{R}^{N} d{\cal S}_{R}, ~.\label{eq4.8}
\end{eqnarray}
where $T_{L}^{N}$ and $T_{R}^{N}$ are the exact left-moving and right-moving temperatures 
of the dual CFT. It was first reported in~\cite{chen12,chen13} for four dimensional KN BH.

Let us assume that the CFT entropies should be reduced by Cardy's 
formula 
\begin{eqnarray}
\mathcal {S}_{L} &=& \frac{\pi^2}{3} c_{L}^{N} T_{L}^{N}~.\label{eq4.9}
\end{eqnarray}
and 
\begin{eqnarray}
\mathcal S_{R} &=& \frac{\pi^2}{3} c_{R}^{N} T_{R}^{N}~.\label{eq5.0}
\end{eqnarray}
from which we can derive the central charges. It should be easily proved that if the entropy 
product function is mass-independent then the left-moving and right-moving sector central 
charges must be equal~\cite{chen12} i.e.  
\begin{eqnarray}
c_{L}^{N}=c_{R}^{N} ~\label{eq5.1}
\end{eqnarray}
Putting  ${\cal S}_{+}={\cal S}_{L}+{\cal S}_{R}$ and ${\cal S}_{-}={\cal S}_{L}-{\cal S}_{R}$
in Eq.~(\ref{e1.1}) and take variations on both sides of the equations while keeping $J_{N}$ is 
constant, and we obtain 
\begin{eqnarray}
\left(\frac{\partial {\cal F}}{\partial N} \right) dN &=& 
\frac{{\cal S}_{L}\,d{\cal S}_{L}-{\cal S}_{R}\,d{\cal S}_{R}}{2\pi^2} ~\label{eq5.2}
\end{eqnarray}
Taking the advantage from Eq.~(\ref{eq4.9}), Eq.~(\ref{eq5.0}), Eq.~(\ref{eq5.1}) and Eq.~(\ref{eq4.8}), 
we get  the central charges 
\begin{eqnarray}
c_{L}^{N} &=& 6  \left(\frac{\partial {\cal F}}{\partial N} \right)=24N^3 \\
c_{R}^{N} &=& 6  \left(\frac{\partial {\cal F}}{\partial N} \right)=24N^3 ~\label{eq5.3}
\end{eqnarray}
These are the central charges of dual CFT in the  NUT picture and we proved that they are equal. 
Similarly, we should keep $N$ is constant and find the $J_{N}$ picture  in which the dual CFT 
is that of the central charges 
\begin{eqnarray}
c_{L}^{J_{N}} &=& 6  \left(\frac{\partial {\cal F}}{\partial J_{N}} \right)=12 J_{N}\\
c_{R}^{J_{N}} &=& 6  \left(\frac{\partial {\cal F}}{\partial J_{N}} \right)=12 J_{N} ~\label{eq5.4}
\end{eqnarray}
From the above analysis, remarkably we obtain a relation between the EPF 
and the central charges  as 
\begin{eqnarray}
c_{L}^{i} &=& 6  \left(\frac{\partial F}{\partial {\cal N}_{i}} \right)\\
c_{R}^{i} &=& 6  \left(\frac{\partial F}{\partial {\cal N}_{i}} \right) ~\label{eq5.5}
\end{eqnarray}
This is the master formula of this work. Where ${\cal N}_{i}$ is one of the integer-valued 
charges appering in the first laws. It 
should be a NUT charges or another new conserved charges i.e. $J_{N}$. Thus 
\begin{eqnarray}
c_{L}^{i} &=& c_{R}^{i}  ~\label{q5.5}
\end{eqnarray}
It must be mentioned that the Eq.~(\ref{eq4.6}) indicates how the BH responds to various 
types of perturbations. If we could consider the perturbations carrying only NUT charges, 
or more precisely $(dN, dJ_{N})=dN(1,0)$, the corresponding first law [Eq.~\ref{eq4.8}] gives the 
exact left-moving and right-moving temperatures,  $T_{L}^{N}$ and $T_{R}^{N}$ of dual 
CFT and then the central chrages $c_{L,R}^{N}$ in the NUT picture. 

On the other side, if we consider the perturbations carrying only new conserved charges, $J_{N}$
or more precisely $(dN, dJ_{N})=dJ_{N}(0,1)$, the first law gives the exact left-moving and right-moving 
temperatures,  $T_{L}^{J_{N}}$ and $T_{R}^{J_{N}}$ of dual CFT in the $J_{N}$ picture and then the central 
chrages $c_{L,R}^{J_{N}}$. 

Now if we consider the perturbations simultaneously i.e. $(dN, dJ_{N})=d{\cal N}(a,b)$ with $a$, $b$ 
being two coprime integers. Then from the first laws we could write 
\begin{eqnarray}
\frac{dM}{2} &=& T_{L}\, d{\cal S}_{L}+\left(a\,\omega_{L}\, dJ_{N}+b\,\psi_{L}\, dN\right)\nonumber \\
             &=& T_{R}\, d{\cal S}_{R}+\left(a\,\omega_{R}\, dJ_{N}+b\,\psi_{R}\, dN\right)  ~.\label{q4.6}
\end{eqnarray} 
Using a similar procedure we get the dual picture i.e. $(a,b)$ picture, in which the CFT is that of the central 
charges
\begin{eqnarray}
 c_{L}^{(a,b)} &=& a\, c_{L}^{N}+b\, c_{L}^{J_{N}}\\
 c_{R}^{(a,b)} &=& a\, c_{R}^{N}+b\, c_{R}^{J_{N}}
\end{eqnarray}
Now we will introduce the B\'{e}zout's identity which states that every pair of coprime integers $a$, $b$ 
there exist other pairs of coprime integers $c$, $d$ such that $ad-bc=1$. Thus the $(a,b)$ picture should 
be viewed as being generated from two elementary (1,0) and (0,1) pictures by a $SL(2,Z)$ transformation:
\begin{eqnarray}
\left( \begin{array}{c} c_{L,R}^{(a,b)} \\ c_{L,R}^{(c,d)} \end{array} \right)
=\left(\begin{array}{cc} a & b \\ c & d \end{array} \right)
\left( \begin{array}{c} c_{L,R}^{(1,0)} \\ c_{L,R}^{(0,1)} \end{array} \right), ~~~
\left(\begin{array}{cc} a & b \\ c & d \end{array} \right) \in SL(2,Z). \nonumber
\end{eqnarray}
So for NUT class of BH, we can generate more holographic descriptions labeled by a pair of 
coprime integers $(a,b)$. These pictures are constructed on two elementary pictures and related 
to each other by $SL(2,Z)$ duality.  More  holographic pictures indicates towards multiple
holographic duals which play a key role to understanding the quantum nature of gravity.
%$$$$$$$$$$$$$$$$$$$$$$$$$$$$$$$$$$$$$$$$$$$$$$$$$4
\section{\label{etn}Exact calculation of central charges for TNUT BH}
In the previous section we derived the central charges by using EPF.
In the present section we will reverify the above result by direct calculation in both sectors. 
To derive the microscopic entropy via the Cardy formula we have to calculate the following 
important thermodynamic parameters in left-moving sectors and right-moving sectors:
\begin{eqnarray}
T_{L} &=& \frac{1}{4\pi \left(r_{+}+ r_{-} \right)}, \,\,\, 
T_{R} = \frac{1}{4\pi \left(r_{+}- r_{-} \right)} \nonumber\\
\mathcal {S}_{L} &=&  \frac{\pi (r_{+}-r_{-})^2}{2} , \,\,\, 
\mathcal {S}_{R} =  \frac{\pi (r_{+}^2-r_{-}^2)}{2} \nonumber\\
\omega_{L} &=& 0,  \,\,\, \omega_{R} =\frac{N}{(r_{+}-r_{-})^2}\nonumber\\
\Psi_{L} &=& -\frac{N}{r_{+}-r_{-}}  , \,\,\, \Psi_{R} =-\frac{N(r_{+}+r_{-})}{(r_{+}-r_{-})^2}   
~.\label{eq6.1}
\end{eqnarray}
There exists two holographic picture for TNUT BH. They are defined as $N$ picture and $J_{N}$ 
picture. Now we have to calculate the dimensionless temperature of the left and right
moving sectors of the dual CFT in $J_{N}$ picture. It could be found as 
\begin{eqnarray}
T_{L}^{J_{N}} &=& \frac{1}{4\pi N} \frac{\left(r_{+}-r_{-} \right)^2}{\left(r_{+}+r_{-} \right)}~.\label{eq6.2}
\end{eqnarray}
\&
\begin{eqnarray}
T_{R}^{J_{N}} &=& \frac{1}{4\pi N} \left(r_{+}-r_{-} \right)~.\label{eq6.3}
\end{eqnarray}
These are exactly the microscopic temperature of dual CFT in TNUT spacetime.

Now we are ready to determine the central charges in left and right moving sectors of the TNUT/CFT 
correspondence via the Cardy formula 
\begin{eqnarray}
\mathcal {S}_{L}^{J_{N}} &=& \frac{\pi^2}{3}c_{L}^{J_{N}}T_{L}^{J_{N}},\,\,\, \mathcal {S}_{R}^{J_{N}} = \frac{\pi^2}{3}c_{R}^{J_{N}}T_{R}^{J_{N}}
~.\label{eq6.4}
\end{eqnarray}
Hence the central charges of dual CFT  becomes
\begin{eqnarray}
c_{L}^{J_{N}} &=& 12J_{N},\,\,\,  c_{R}^{J_{N}}=12J_{N}    ~.\label{eq6.5}
\end{eqnarray}
It implies that the central charges of left moving sectors and right moving sectors of 
dual CFT are same for TNUT BH. This is a remarkable result for \emph{TNUT BH}. This is 
possible only due to introduction of new conserved charges i.e. $J_{N}$. The result is 
exactly same as we have seen in case of  Kerr BH~\cite{hartman9} and KN BH~\cite{chen12}. 
This kind of observation tells us that TNUT BH is dual to  
$c_{L}^{J_{N}}=c_{R}^{J_{N}}=12J_{N}$ of 2D CFT at temperature 
$(T_{L}^{J_{N}},T_{R}^{J_{N}})$ for each value of $M$ and $J_{N}$.

Analogously, for $N$ picture, the dimensionless temperature of the left and right
moving sectors of the dual CFT are
\begin{eqnarray}
T_{L}^{N} &=& \frac{\left(r_{+}-r_{-} \right)^2}{16\pi N^3}~.\label{eq6.6}
\end{eqnarray}
and
\begin{eqnarray}
T_{R}^{N} &=& \frac{ \left(r_{+}^2-r_{-}^2\right)}{16 \pi N^3}~.\label{eq6.7}
\end{eqnarray}
These are exactly the microscopic temperature of dual CFT in $N$ picture for 
TNUT spacetime. Therefore the central charges of dual CFT are computed to be 
\begin{eqnarray}
c_{L}^{N} &=& 24 N^3,\,\,\,  c_{R}^{N}=24 N^3    ~.\label{eq6.8}
\end{eqnarray}
These results are completely agreement with the result obtained in 
previous section. With the above thermodynamic parameters one can easily 
check that for TNUT BH
\begin{eqnarray}
\frac{\mathcal {S}_{L}}{\mathcal {S}_{R}}=\frac{T_{L}}{T_{R}} 
\end{eqnarray}
which is definitely satisfied. Where $\mathcal {S}_{L,R}$ are the 
left moving entropies  and right moving entropies of the dual CFT, and 
$T_{L,R}$ are the CFT temperatures.
Now we will give more examples for derivations of central charges 
by using \emph{EPF}. To do that we have to consider first 
RN-NUT BH.
%%%%%%%%%%%%%%%%%%%%%%%%%%%%%%%%%%%%%%%%%%%%%%%%%%%%%%%%%%%%%%
\section{\label{rnn} Central charges for RN-NUT BH from EPF }
For RN-NUT BH,  the metric function ${\cal A}$  should be
\begin{eqnarray}
 {\cal A} &=& \frac{1}{r^2+N^2} \left(r^2-N^2-2Mr+Q^2 \right)
\end{eqnarray}
where $Q$ is the purely electric charge. Now the BH horizons are located at 
\begin{eqnarray}
r_{\pm}= M \pm\sqrt{ M^2-Q^2+N^2}
\end{eqnarray}
Now the entropy~\cite{plb20,grg21} of ${\cal H}^{\pm}$  
is computed under the criterion stated in Eq.~(\ref{ptn})
\begin{eqnarray}
{\cal S}_{\pm} &=& \pi\left[2 \left(M^2+N^2 \right)-Q^2 \pm 2\sqrt{M^4-M^2Q^2+J_{N}^2}\right] ~.\label{eq7}
\end{eqnarray}
The EPF for this BH is
\begin{eqnarray}
{\cal F} &=& J_{N}^2+\left(\frac{Q^2}{2}-N^2\right)^2 ~\label{eq7.0}
\end{eqnarray}
For $J_{N}$ picture, we get  the central charges for left moving and right-moving sectors  
\begin{eqnarray}
c_{L}^{J_{N}} &=& 6 \left(\frac{\partial {\cal F}}{\partial J_{N}} \right)=12J_{N}\\
c_{R}^{J_{N}} &=& 6 \left(\frac{\partial {\cal F}}{\partial J_{N}} \right)=12J_{N}  ~\label{eq7.1}
\end{eqnarray}
It proves that central charges are equal in both sectors i.e. 
\begin{eqnarray}
c_{L}^{J_{N}} &=& c_{R}^{J_{N}} ~\label{e7.1}
\end{eqnarray}
Similarly for $Q$ picture, we get the equal central charges for left moving and right-moving sectors 
\begin{eqnarray}
 c_{L}^{Q} &=& 12Q \left(\frac{Q^2}{2}-N^2\right)\\
 c_{R}^{Q} &=& 12Q \left(\frac{Q^2}{2}-N^2\right)~\label{eq7.2}
\end{eqnarray}
Analogously for $N$ picture, the equal central charges for left moving and right-moving sectors are 
\begin{eqnarray}
 c_{L}^{N} &=& 24N \left(N^2-\frac{Q^2}{2}\right)\\
 c_{R}^{N} &=& 24N \left(N^2-\frac{Q^2}{2}\right)~\label{eq7.3}
\end{eqnarray}
Alternatively, we could say that the central charges are equal means the EPF is 
universal (mass-independent).

\section{\label{knn} Central charges for KNUT BH from EPF}
The metric of KNUT BH is
\begin{eqnarray}
ds^2 &=& -\frac{\Delta_{r}}{\rho^2} \, \left[dt-X d\phi \right]^2+\frac{\sin^2\theta}{\rho^2}\,
\left[(r^2+a^2+N^2)\,d\phi-a\,dt\right]^2+\rho^2 \, \left[\frac{dr^2}{\Delta_{r}}+d\theta^2\right]
~.\label{eq8}
\end{eqnarray}
where
\begin{eqnarray}
a &\equiv&\frac{J}{M},\, \rho^2 \equiv r^2+(N+a\,\cos\theta)^2 \\
\Delta_{r} &\equiv& r^2-2Mr+a^2-N^2\\
X & \equiv& a\,\sin^2\theta-2N\,\cos\theta  ~.\label{ad}
\end{eqnarray}
and  the global conserved charges  are the Komar mass $M$, angular momentum $J=a M$ and 
gravitomagnetic charge or dual mass or NUT parameter $N$. 

The horizons are located at
\begin{eqnarray}
r_{\pm} &=&  M \pm\sqrt{ M^2-a^2+N^2}
\end{eqnarray}
Introducing $J_{N}=MN$, the entropy of ${\cal H}^\pm$ should be  derived to be \cite{plb20,grg21}
\begin{eqnarray}
{\cal S}_{\pm} &=& 2\pi\left[(M^2+N^2) \pm  \sqrt{M^4+J_{N}^2-J^2} \right] ~.\label{8.1}
\end{eqnarray}
The EPF for KNUT BH is defined as
\begin{eqnarray}
{\cal F} &=& J^2+J_{N}^2+N^4 ~\label{eq8.2}
\end{eqnarray}
Hence for $J$ picture, we find  the equal central charges for left moving and right-moving sectors  
\begin{eqnarray}
c_{L}^{J} &=& 12J\\
c_{R}^{J} &=& 12J~\label{eq8.3}
\end{eqnarray}
Similarly for  $J_{N}$ picture, we find the equal  central charges for left moving and right-moving sectors 
\begin{eqnarray}
c_{L}^{J_{N}}=c_{R}^{J_{N}}=12J_{N}~\label{eq8.4}
\end{eqnarray}
Analogously for $N$ picture, the equal central charges for left moving and right-moving sectors are 
\begin{eqnarray}
c_{L}^{N}=c_{R}^{N}=24N^3~\label{eq8.5}
\end{eqnarray}

\section{\label{knnn} Central charges for KNNUT BH from EPF}
The metric for KNNUT BH can be written as
\begin{eqnarray}
ds^2 &=& -\frac{\Delta_{r}}{\rho^2} \, \left[dt-X d\phi \right]^2+\frac{\sin^2\theta}{\rho^2}\,
\left[(r^2+a^2+N^2)\,d\phi-a\,dt\right]^2+\rho^2 \, \left[\frac{dr^2}{\Delta_{r}}+d\theta^2\right]
~.\label{eq9}
\end{eqnarray}
where
\begin{eqnarray}
a &\equiv&\frac{J}{M},\, \rho^2 \equiv r^2+(N+a\,\cos\theta)^2 \\
\Delta_{r} &\equiv& r^2-2Mr+a^2+Q^2-N^2\\
X & \equiv& a\,\sin^2\theta-2N\,\cos\theta  ~.\label{9.1}
\end{eqnarray}
and here the conserved charges are Komar mass $M$, angular momentum $J=a M$, 
gravito-electric charge $Q$ and the gravitomagnetic charge or dual~(magnetic) mass 
or NUT charge $N$. The BH horizons are 
\begin{eqnarray}
r_{\pm} &\equiv &  M \pm \sqrt{M^2-a^2-Q^2+N^2}
\end{eqnarray}
Incorporating $J_{N}=MN$, the entropy of ${\cal H}^\pm$ should be derived as  
~\cite{plb20,grg21}
\begin{eqnarray}
{\cal S}_{\pm} &=& 2\pi\left[2(M^2+N^2)-Q^2 \pm 2 \sqrt{M^4+J_{N}^2-J^2-M^2Q^2} \right]~.\label{9.2}
\end{eqnarray}
The EPF for KNNUT BH is defined to be
\begin{eqnarray}
{\cal F} &=& \left[J^2+J_{N}^2+\left(\frac{Q^2}{2}-N^2\right)^2\right]  ~\label{eq9.3}
\end{eqnarray}
Hence we find  the equal central charges for left moving and right-moving sectors  for $J$ picture
\begin{eqnarray}
c_{L}^{J}=c_{R}^{J}=12J=12aM=12a\sqrt{a^2+Q^2-N^2}~\label{9.4}
\end{eqnarray}
This central charges in $J$ picture are completely agreement with the central charges  derived 
by using Brown-Henneaux technique which makes the use of asymptotic symmetry group~(ASG)~
\cite{sakti18}[Shortly derived in Appendix C].
Similarly we find the equal central charges for left moving and right-moving sectors  for  $J_{N}$ picture
\begin{eqnarray}
c_{L}^{J_{N}}=c_{R}^{J_{N}}=12J_{N}~\label{eq9.4}
\end{eqnarray}
Analogously the equal central charges for left moving and right-moving sectors are  in $N$ picture,
\begin{eqnarray}
c_{L}^{N}=c_{R}^{N}=24N \left( N^2-\frac{Q^2}{2}\right)~\label{eq9.5}
\end{eqnarray}
Finally for $Q$ picture, we get the equal central charges for left moving and right-moving sectors 
\begin{eqnarray}
 c_{L}^{Q}=c_{R}^{Q}=12Q \left(\frac{Q^2}{2}-N^2\right)~\label{eq9.6}
\end{eqnarray}
The above analysis tells us that using EPF one could derive central charges 
for various pictures of conserved charges. It also remarkable that the central 
charges are equal means the \emph{EPF is mass-independent}. On the other hand 
the central charges are unequal means the \emph{EPF is mass-dependent}. 
This is an another way one could verify whether the entropy product 
of ${\cal H}^{\pm}$ is mass-independent or not.

\section{\label{con} Conclusions}
It was proposed that a generic four dimensional TNUT BH could be explicitly expressed 
in terms of three or four different types of thermodynamic hairs. They must be defined 
as the Komar mass~($M=m$), the angular momentum~($J_{n}=mn$), the gravitomagnetic 
charge~($N=n$), and or the dual~(magnetic) mass $(\tilde{M}=n)$. In this context, we 
defined  the EPF for a TNUT BH. Using this feature and the above proposal, we evaluated  
the \emph{central charges} of dual CFT by virtue of Cardy's formula. 

Remarkably for NUT class of BH,  we proved that there exists an 
\emph{universal relation between the central charges and EPF} 
as $c=6\left(\frac{\partial {\cal F}}{\partial {\cal N}_{i}}\right)$. We first showed the entropy 
product function could be derived by using first law of thermodynamics of both the horizons.
Also, we computed the different thermodynamic  parameters in the left-moving sectors and 
right-moving sectors and reverified the results obtained by using EPF. 
%We showed that the result is completely agreement with the previous one. 
Moreover, we examined the first laws that satisfied for both the horizons 
in case of left-moving and right-moving sectors.

Furthermore, we introduced the B\'{e}zout's identity and showed that for NUT 
class of BH we can generate more holographic pictures that described by a pair 
of intgers $(a,b)$. Again more holographic picture of NUT class of BH indicates 
toward understanding the holographic  nature of quantum gravity. At the end we 
showed  that using EPF one could derive central charges for various NUT class 
of BHs. Remarkably, we found that the central charges are equal in both sectors 
and this immediately indicates that the \emph{EPF is mass-independent}. 
This is an another way one could verify whether the entropy product is 
mass-independent or not. Finally, we showed that the EPF method is more 
convenient to derive the central charges than the ASG method.

\section{Appendix-A}
In this appendix section, we will calculate the central charges from EPF 
for four dimensional RN BH, Kerr BH and KN BH and contrast with the central charges 
derived from the near-horizon geometry of the corresponding extremal BH by using the  
asymptotic symmetry group~(ASG) introduced by Brown and Henneaux~\cite{brown}.

1) \emph{Kerr BH}\\
The EPF for Kerr BH is 
\begin{eqnarray}
{\cal F} &=& J^2 ~\label{apd1.3}
\end{eqnarray}
The equal central charges for left moving and right-moving sectors obtained for $J$ picture as
\begin{eqnarray}
c_{L}^{J} &=& 12 J\\
c_{R}^{J} &=& 12J~\label{app1.4}
\end{eqnarray}
This result is completely agreement with the central charges obtained by ASG analysis~\cite{guica}.

2)\emph{KN BH}\\
The EPF for KN BH is given by
\begin{eqnarray}
{\cal F} &=& J^2+\frac{Q^4}{4} ~\label{apd1.4}
\end{eqnarray}
Hence  for $J$ picture we find  the central charges for left moving and right-moving sectors are 
\begin{eqnarray}
c_{L}^{J} &=& 12J ~\\
c_{R}^{J} &=& 12J  \label{app1.5}
\end{eqnarray}
These results are completely agreement with the central charges obtained by ASG analysis in Eq.~(\ref{apd2.2}).
Similarly for $Q$ picture, one finds the equal central charges for left moving and right-moving sectors 
\begin{eqnarray}
c_{L}^{Q}&=&6Q^3 \\
c_{R}^{Q}&=& 6Q^3~\label{apd1.6}
\end{eqnarray}
Similarly, these results are also completely agreement with the central charges obtained by ASG 
analysis~\cite{hartman9}.

\section{Appendix B: Calculation of Central charges by using Near-Horizon Geometry of Extremal KN BH}
In this appendix section, we will review how to derive central charges of extremal KN BH 
in the near-horizon limit by using ASG analysis followed by the work of Hartman~\cite{hartman9}. 
To do that we have to consider near horizon metric of an extremal, stationary, rotationally 
symmetric BH of the following form
\begin{eqnarray}
 ds^2 &=& \Upsilon \left[-r^2\, dt^2+\frac{dr^2}{r^2}+\zeta\, d\theta^2 \right]
 +\chi \left(d\phi+kr\,dt \right)^2~\label{apd2.1}
\end{eqnarray}
where
$$
\Upsilon=\rho_{+}^{2},\, \zeta=1,\, \chi=\frac{(r_{+}^2+a^2)^2\sin^2\theta}{\rho_{+}^2}
$$
and 
we have defined 
$$
\rho_{+}^2=r_{+}^2+a^2\cos^2\theta,\, k=\frac{2ar_{+}}{r_{+}^2+a^2}, \, r_{+}=m
$$
Then the central charges is derived to be in the $J$ picture
\begin{eqnarray}
c_{L} &=& 3k\int_{0}^{\pi}\sqrt{\Upsilon\,\zeta\, \chi}\,d\theta\\
      &=& 12ar_{+}=12am=12J=12a\sqrt{a^2+Q^2} ~\label{apd2.2}
\end{eqnarray}

\section{Appendix C: Calculation of Central charges by using Near-Horizon Geometry of Extremal KN-NUT BH}
In this appendix section, we will review how to derive central charges of extremal KN-NUT BH 
in the near-horizon limit by using ASG analysis followed by the work~\cite{sakti18}. 
To proceed we have to consider near horizon metric of an extremal, stationary, rotationally symmetric 
BH of the following form
\begin{eqnarray}
 ds^2 &=& \Upsilon \left[-y^2\, d\tau^2+\frac{dy^2}{y^2}+\zeta\, d\theta^2 \right]
 +\chi \left(d\phi+p\,y\,d\tau \right)^2~\label{c1}
\end{eqnarray}
where
$$
\Upsilon=\rho_{+}^{2},\, \zeta=1,\, \chi=\frac{[r_{+}^2+(a+n)^2]^2\sin^2\theta}{\rho_{+}^2}
$$
and 
we have defined 
$$
\rho_{+}^2=r_{+}^2+(n+a\cos\theta)^2,\, p=\frac{2ar_{+}}{r_{+}^2+(a+n)^2}, \, r_{+}=m
$$
Then the central charges is derived to be in the $J$ picture
\begin{eqnarray}
c_{L} &=& 3p\int_{0}^{\pi}\sqrt{\Upsilon\,\zeta\, \chi}\,d\theta\\
      &=& 12ar_{+}=12am=12J ~\label{c2}
\end{eqnarray}

\end{document}